\documentclass[12pt, oneside, a4paper, final]{article}

\usepackage[version=3]{mhchem} %Escrever corretamente as formulas
\usepackage[scriptsize,tight]{subfigure}
\usepackage{graphicx}
\usepackage{multirow}
\usepackage{dcolumn}% Align table columns on decimal point
\usepackage{bm}% bold math
\usepackage{color,soul}
\usepackage[mathlines]{lineno}
\usepackage{hyperref}
\usepackage[text={16cm,23cm},centering]{geometry}
\usepackage{listings}

\definecolor{dkgreen}{rgb}{0,0.6,0}
\definecolor{gray}{rgb}{0.5,0.5,0.5}
\definecolor{mauve}{rgb}{0.58,0,0.82}

\begin{document}

%\setpagewiselinenumbers

%\modulolinenumbers[5]
%\linenumbers

\title{Methods used in nanostructure modeling}

\footnotesize\date{\today}

\author{I. Camps\\
\footnotesize Laborat\'orio de Modelagem Computacional - \emph{La}Model, \\
\footnotesize Instituto de Ci\^{e}ncias Exatas - ICEx. Universidade Federal de
\footnotesize Alfenas - UNIFAL-MG, \\
\footnotesize Alfenas, Minas Gerais, Brasil\\
\footnotesize HPQC Labs, Waterloo, Canada
\footnotesize \texttt{icamps@unifal-mg.edu.br} \\ }

\maketitle

\begin{abstract}

How many times you need to change your method description because you were
``accused'' of plagiarism from text you already published? I will use this
preprint to add all the methods I currently used in running the simulations for
my research works. Then, I will cite it as needed.

\end{abstract}

%\tableofcontents

\section{Functionalizing Structures}
In case of multiple functionalizations, populations of 10000 structures were
generated for each system with the functional group (-OH, -COOH, etc.) added
randomly to the surface sites. To choose a representative structure for each
group/concentration, first, using OpenBabel software~\cite{Openbabel}, the
extended-connectivity fingerprints (ECFP) with bond diameter four
(ECFP4)~\cite{ECFP4} was generated for each structure. ECFP4 was selected as it
is currently one of the most used for similarity
searching~\cite{maggiora-J.Med.Chem.-57-3186}. Then, the
entropy of the binary fingerprint was calculated using the
BiEntropy function~\cite{BiEntropy}. This function is capable to identify order
and disorder in binary strings. Finally, the most common structure was selected
from the entropy distribution. Due to the volume of some groups (like -COOH)
and the size of the base system used, it is only possible to add some
percentage (5\%, 10\%,  15\%, 20\% and/or 25\%) to cover the surface sites.

\section{Metal nanoclusters}
\begin{figure}[tbph]
\centering
\begin{tabular}{ccc}
\subfigure[1DL]{\includegraphics[width=3.5cm]{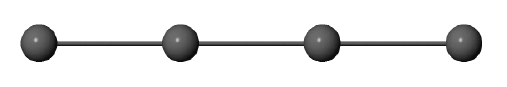}
\label{Fig:1DL}}      &
\subfigure[1DZ]{\includegraphics[width=3.5cm]{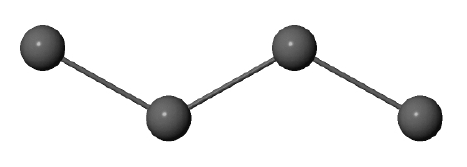} \label{Fig:1DZ}} \\
\subfigure[2D]{\includegraphics[width=3.5cm]{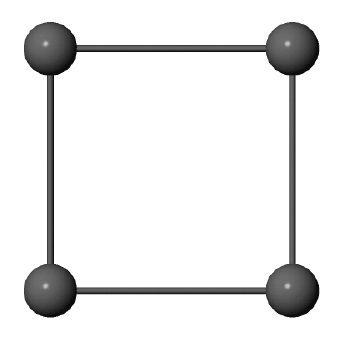} \label{Fig:2D}}
&  \subfigure[3D]{\includegraphics[width=3.5cm]{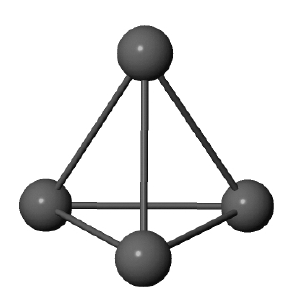} \label{Fig:3D}}
\end{tabular}
\caption{\label{Fig:Nanoclusters} Metal nanoclusters geometry: (a) one
dimensional
linear chain,
(b) one dimensional zigzag chain, (c) two dimensional plane, and (d) three
dimensional tetrahedron.}
\end{figure}

The metals considered here are in the form of isolated atoms and forming
four--atom nano\-clus\-ters. In case of nanoclusters, were considered four
different configurations: one dimensional linear chain (1DL,
figure~\ref{Fig:1DL}), one dimensional zigzag chain (1DZ,
figure~\ref{Fig:1DZ}), two dimensional plane (2D, figure~\ref{Fig:2D}) and
three dimensional tetrahedron (3D, figure~\ref{Fig:3D}). For notation
completeness, the isolated atoms are labeled as 0D.

A script to generate the metal nanocluster is given in
section~\ref{script_nanoclusters}.

\section{SIESTA}
All the studies presented in this work were done using the density functional
theory (DFT)~\cite{Martin2020}. We used the SIESTA software~\cite{siesta} to
optimize the geometry of the periodic structures and to calculate the
electronic bands, density of states and binding energy. In this case, the
generalized gradient approximation (GGA) was employed for the
exchange-correlation potential using the PBE
scheme~\cite{perdew-prl-77-3865}. The valence electrons were treated using a
split-valence double-$\zeta$ basis set with polarization functions
(DZP)~\cite{artaho-pssb-215-809} whereas the core electrons were represented
using the norm-conserving Troullier-Martins pseudo
potentials~\cite{troullier-prb-43-1991}. The accuracy of the results was
guaranteed via convergence studies on the mesh-cutoff energy and the number
of \textbf{k}-points. The energy convergence for the systems was obtained for
a mesh-cutoff of $500$~Ry.  For the number of \textbf{k}-points, the
Brillouin zone was sampled in accordance with Monkhorst and
Pack~\cite{monkhorst-prb-13-5188} with a $1\times1\times20$ \textbf{k}-point
sampling set. All the properties were calculated on systems with fully
relaxed atomic coordinates during geometry optimization until the
Hellman-Feynman forces were below $0.03$~eV\AA$^{-1}$.

The atomic charge distribution was also determined from first principle
calculations. In this work, we examine the charge density distribution using
the Hirshfeld partition scheme~\cite{HirshfeldCharges} implemented in the
SIESTA software. This partition scheme has proven to be highly insensitive to
the choice of the basis set~\cite{HirshfeldCharges,f.martin-jcc-26-97}.

\section{Charges}
One of the properties that can be calculated within the quantum mechanical
approach is the partial atomic charge distribution. To accomplish this task,
there are several methodologies like (i) Mulliken population
analysis~\cite{mulliken1,mulliken2,mulliken3,mulliken4}, (ii) natural
population analysis (NPA)~\cite{charges_NPA}, (iii) the Breneman-Wiberg (BW)
model~\cite{charges_BW}, (iv) Merz-Kollman-Singh (MKS) electrostatic
potential derived charges~\cite{charges_mks1,charges_mks2}, (v) the Bader
partition scheme~\cite{Bader1994,bader1,bader2,bader3}, (vi) the Voronoi
partition scheme~\cite{voronoi1,voronoi2} and (vii) the Hirshfeld partition
scheme~\cite{HirshfeldCharges}. (vii) the CM5
scheme~\cite{charges_CM5}. However, none of them is universally
accepted
as the ``best" for computing partial atomic charges.

\section{xTB}
To perform the calculations, we used the semiempirical tight binding method, as
implemented in the xTB program~\cite{xTB_1,xTB_2}. Within the xTB program,
there are implemented three different methods called GFNn-xTB, with $n=0, 1,
2$. GFN0-xTB~\cite{xTB_GFN0}  is parameterized for all elements up to radon and
includes only quantum mechanical contributions up to first--order, a classical
electronegativity equilibration charge model and the atomic charge-dependent D4
dispersion correction to account for long range London correlation effects.
GFN1-xTB uses similar second order with some
terms up to third order
approximations for the Hamiltonian and electrostatic energy as
DFTB3~\cite{Gaus_2011}. GFN2-xTB~\cite{xTB_GFN2} uses a multipole electrostatic
treatment up to
quadrupole terms and the latest D4 dispersion model~\cite{Caldeweyher_2019}

\subsection{Docking}
Using the automated Interaction Site
Screening (aISS)~\cite{xTB-dock}, we generated different intermolecular
geometries of the complexes and subjected them to further genetic
optimization. The main idea of aISS is to find the global energy minimum
(energetically lowest structure) of the largest interaction between two
molecules. Firstly, a step is run to search for pockets in
molecule A and then, a screening for $\pi-\pi$~--stacking interactions along
different directions in three dimensions (3D). The next step is a search for
global orientations of molecule B on an angular grid around molecule A. We used
the interaction energy (xTB-IFF)~\cite{xTB-IFF} of each
newly generated structure for ranking, and the genetic step was repeated ten
times until the best complex was obtained. The best--ranked complexes were
further subjected to structural optimization.

\subsection{Geometry optimization}
All geometry optimizations were performed using the GFN2--xTB method, which is
an accurate self--consistent method that includes multipole electrostatics and
density--dependent dispersion contributions~\cite{xTB_GFN2}. Extreme
optimization level was ensured, with a convergence energy of
$5\times10^{-8}$~E\textsubscript{h} and gradient norm
convergence of $5\times10^{-5}$~E\textsubscript{h}/a\textsubscript{0} (where
a\textsubscript{0} is the Bohr radius).

\section{Topological analysis}
On topological analysis, one of the objectives is to determine the critical
points. From the topological analysis point of view, critical points (CPs)
are points where the gradient norm of the function value is zero. They are
classified into four types depending on how the eigenvalues of Hessian matrix
of real function are negative~\cite{Bader1994}.

The CPs identified as $\bf (3,-3)$ occurs when three eigenvalues of the
Hessian matrix are negative. They positions are nearly identical to atomic
positions thus they are called nuclear critical point (NCP).

If two eigenvalues of the Hessian are negative (second--order saddle point),
they are represented as $\bf (3,-1)$. For electron density analysis, they
generally appear between atom pairs, thus they are called bond critical point
(BCP). The value of the electron density ($\rho$) and the sign of its
Laplacian ($\nabla ^2\rho$) can be related to the strength and type of the
bonds formed~\cite{Matta2007}.

If one eigenvalue of the Hessian is negative (first--order saddle point),
they are represented as $\bf (3,+1)$. For electron density analysis, they
generally appear in the center of a ring system, thus they are called ring
critical point (RCP).

Finally, if none of the eigenvalues are negative (local minimum) they are
represented as $\bf (3,+3)$. For electron density analysis, they generally
appear in the center of a cage system, thus they are called cage critical
point (CCP).

One way to classify the strength of bond (covalent or non-covalent) is to
look at the electron density ($\rho$) and the sign of its Laplacian ($\nabla
^2\rho$). Values of $\rho>0.20\,a.u.$ indicate a covalent bond whereas
$\rho<0.10\,a.u.$ indicate a non-covalent bond. On the other hand, if $\nabla
^2\rho<0$, the bond can be classified as covalent and if $\nabla ^2\rho>0$
can be classified as non covalent~\cite{Matta2007}. The ELF index is related
to the electron movement confinement. Its value is in the range of
$\left[0,1\right]$. Large values mean that electrons are greatly localized
indicating the presence of a covalent bond. The LOL index is another function
for locating high localized regions~\cite{lol}. Values of LOL index are also
in the range of $\left[0,1\right]$. Smaller (large) values usually appear in
boundary (inner) regions.

\section{Conclusion}
We described here the main methods we currently use when running our computer
simulations during our research.

%\bibliographystyle{elsarticle-num}
%\bibliography{unifal}

\pagebreak \clearpage
\begin{center}
\textbf{\large Supplemental Materials: Methods used in nanostructure modeling}
\end{center}

%%%%%%%%%% Merge with supplemental materials %%%%%%%%%%
\setcounter{section}{0}
%%%%%%%%%% Prefix a "S" to all equations, figures, tables and reset the counter
%%%%%%%%%%%%%%%%%%%%%%%%%%%%%%
\setcounter{equation}{0} \setcounter{figure}{0} \setcounter{table}{0}
\setcounter{page}{1} \makeatletter
\renewcommand{\thesection}{S\arabic{section}}
\renewcommand{\theequation}{S\arabic{equation}}
\renewcommand{\thefigure}{S\arabic{figure}}
\newcommand{\bibnumfmt}[1]{[S#1]}
\newcommand{\citenumfont}[1]{S#1}
%%%%%%%%%% Prefix a "S" to all equations, figures, tables and reset the counter
%%%%%%%%%%%%%%%%%%%%%%%%%%%%%%

\section{Script to generate the nanoclusters}
\label{script_nanoclusters}
\begin{lstlisting}
#!/bin/bash
# Generate 4 nanoclusters of 4 atoms each.
# 1Dlinear
# 1Dzigzag
# 2D
# 3D
# Ni 2.5
# Cd 2.98
# Pb 3.5
if [ -z $1 ] || [ -z $2 ]
then
    echo "Add the Element and bond distance: gera_nanocluster Ni 2.5    "
else
# arg 1: Atom symbol
# arg 2: Bond distance used
Element=$1
sen60=0.5
cos60=0.866025
# a=`echo "$a*$bohr" | bc`
# 1Dlinear
x1=0
y1=0
z1=0
x2=0
y2=0
z2=$2
x3=0
y3=0
z3=`echo "2*$2" | bc`
x4=0
y4=0
z4=`echo "3*$2" | bc`
cat > ${Element}M4_1Dlinear.xyz <<!
4
${Element}M4_1Dlinear
${Element}   $x1   $y1   $z1
${Element}   $x2   $y2   $z2
${Element}   $x3   $y3   $z3
${Element}   $x4   $y4   $z4
!
# 1Dzigzag
x1=0
y1=0
z1=0
x2=0
y2=`echo "$sen60*$2" | bc`
z2=`echo "$cos60*$2" | bc`
x3=0
y3=`echo "$y2-$sen60*$2" | bc`
z3=`echo "$z2+$cos60*$2" | bc`
x4=0
y4=`echo "$y3+$sen60*$2" | bc`
z4=`echo "$z3+$cos60*$2" | bc`
cat > ${Element}M4_1Dzigzag.xyz <<!
4
${Element}M4_1Dzigzag
${Element}   $x1   $y1   $z1
${Element}   $x2   $y2   $z2
${Element}   $x3   $y3   $z3
${Element}   $x4   $y4   $z4
!
# 2D
x1=0
y1=0
z1=0
x2=0
y2=0
z2=$2
x3=0
y3=$2
z3=0
x4=0
y4=$2
z4=$2
cat > ${Element}M4_2D.xyz <<!
4
${Element}M4_2D
${Element}   $x1   $y1   $z1
${Element}   $x2   $y2   $z2
${Element}   $x3   $y3   $z3
${Element}   $x4   $y4   $z4
!
# 3D
x1=0
y1=0
z1=0
x2=0
y2=0
z2=$2
x3=0
y3=`echo "$cos60*$2" | bc`
z3=`echo "0.5*$2" | bc`
x4=`echo "$cos60*$2" | bc`
y4=`echo "0.5*$y3" | bc`
z4=$z3
cat > ${Element}M4_3D.xyz <<!
4
${Element}M4_3D
${Element}   $x1   $y1   $z1
${Element}   $x2   $y2   $z2
${Element}   $x3   $y3   $z3
${Element}   $x4   $y4   $z4
!
fi
\end{lstlisting}
\section{SLURM script to run calculations using xTB}
\begin{lstlisting}
#!/bin/bash
#SBATCH --mail-user=icamps@gmail.com
#SBATCH --mail-type=ALL
#SBATCH --time=7-0:0
#SBATCH --job-name="BNMobPb3D"
#SBATCH --nodes=1
#SBATCH --ntasks=1
#SBATCH --cpus-per-task=10
#SBATCH --mem=10G
module load StdEnv/2020
#module load xtb/6.5.0
source ~/bin/xtb-bleed/share/xtb/config_env.bash
export MKL_NUM_THREADS=${SLURM_CPUS_PER_TASK}
export OMP_NUM_THREADS=${SLURM_CPUS_PER_TASK},1
export OMP_STACKSIZE=4G
ulimit -s unlimited
filename1="MBNNB"
filename2="PbM4_3D"
T="300"
#level Econv(Eh)/ Gconv (Eh/a0)/ Accuracy
#crude 5E-4/ 1E-2/ 3.00
#sloppy 1E-4/ 6E-3/ 3.00
#loose 5E-5/ 4E-3/ 2.00
#lax 2E-5/ 2E-3/ 2.00
#normal 5E-6/ 1E-3/ 1.00
#tight 1E-6/ 8E-4/ 0.20
#vtight 1E-7/ 2E-4/ 0.05
#extreme 5E-8/ 5E-5/ 0.01
nOpt="extreme"
# --gfn 0
# --gfn 1
# --gfn 2
# --gfnff
hamilton="--gfn 2"
nIter="500" #default 250
# Single molecule geometry optimization
mkdir out_opt
cd out_opt
echo "Optimizing molecule 1..."
mkdir opt_mol_1
cd  opt_mol_1
cat > in_geo-opt.inp <<!
\$scc
    temp=${T}
\$write
    output file=_.out
    esp=true
    density=true
    spin population=true
    spin density=true
    mos=true
    wiberg=true
    charges=true
    mulliken=false
\$opt
    engine=rf
!
xtb ${hamilton} ./../../in_XYZ/${filename1}.xyz --input in_geo-opt.inp --molden
--iterations ${nIter} --opt ${nOpt} -P ${nCPU} --namespace ${filename1} >
opt_${filename1}.log
cd ..
echo "Optimizing molecule 2..."
mkdir opt_mol_2
cd  opt_mol_2
cat > in_geo-opt.inp <<!
\$scc
    temp=${T}
\$write
    output file=_.out
    esp=true
    density=true
    spin population=true
    spin density=true
    mos=true
    wiberg=true
    charges=true
    mulliken=false
\$opt
    engine=rf
!
xtb ${hamilton} ./../../in_XYZ/${filename2}.xyz --input in_geo-opt.inp --molden
--iterations ${nIter} --opt ${nOpt}  -P ${nCPU} --namespace ${filename2} >
opt_${filename2}.log
cd ../../
# DOCKING
mkdir out_dock
cd out_dock
cat > in_dock.inp <<!
\$dock
   pocket
   stack
   maxparent = 100
   nfinal = 10
   atm
\$end
!
echo "Docking..."
Complex=${filename1}+${filename2}
echo ${Complex}
xtb dock ./../in_XYZ/${filename1}.xyz ./../in_XYZ/${filename2}.xyz --input
in_dock.inp --opt ${nOpt} --etemp ${T}> dock.log
dock_filename=dock_$Complex
mv best.xyz ${dock_filename}.xyz
cd ..
# Complex geometry optimization
mkdir out_opt-complx
cd out_opt-complx
echo "Optimizing Complex..."
opt_filename=opt_$Complex
cp ../out_dock/${dock_filename}.xyz .
cat > in_geo-opt.inp <<!
\$scc
    temp=${T}
\$write
    output file=_.out
    esp=true
    density=true
    spin population=true
    spin density=true
    mos=true
    wiberg=true
    charges=true
    mulliken=false
\$opt
    engine=rf
!
xtb ${hamilton} ${dock_filename}.xyz --input in_geo-opt.inp --molden
--iterations ${nIter} --opt ${nOpt} -P ${nCPU} --namespace ${opt_filename} >
opt_Complex.log
cd ..
# Molecular Dynamics
mkdir out_md
cd out_md
md_filename=md_$Complex
cp ../out_dock/${dock_filename}.xyz .
cat > in_md.inp <<!
\$md
   temp=298.15 # in K
   time= 100.0  # in ps
   dump= 50.0  # in fs
   step=  2.0  # in fs
   velo= false
   nve = true
   hmass=4
   shake=2 # constrain all bonds
   sccacc=1.0
\$end
!
echo "Molecular Dynamics..."
echo ${md_filename}
xtb $hamilton ${dock_filename}.xyz --input in_md.inp --md --namespace
${md_filename} -P ${nCPU} --iterations ${nIter} > md.log
mkdir scoord
mv *.scoord* scoord
\end{lstlisting}

\end{document}